\documentclass[15pt]{revtex4}
\usepackage{graphicx}
\usepackage{epstopdf}
\usepackage{dcolumn}
\usepackage{bm}

\begin{document}

\title{Superfluidity of a rarefied gas of electron-hole pairs in a bilayer system}

\author{D. V. Fil}
\affiliation{Institute for Single Crystals National Academy of
Sciences of Ukraine, Nauky Ave. 60, Kharkiv 61001, Ukraine}
\affiliation {V. N. Karazin Kharkiv National University, Svobody Sq.
4, Kharkiv 61022, Ukraine}
\author{S. I. Shevchenko}
\affiliation{B.Verkin Institute for Low Temperature Physics and
Engineering National Academy of Sciences  of Ukraine, Nauky Ave. 47,
Kharkiv 61103, Ukraine}

\begin{abstract}
The conditions of stability of the superfluid phase in double layer
systems with pairing of spatially separated electrons and holes in
the low density limit are studied. The general expression for the
collective excitation spectrum is obtained. It is shown that under
increase in the distance $d$ between the layers the minimum emerges
in the excitation spectrum. When $d$ reaches the critical value the
superfluid state becomes unstable relative to the formation of a
kind of the Wigner crystal state. The same instability occurs at
fixed $d$ under increase in the density of carries. It is
established that the critical distance and the critical density are
related to each other by the inverse power function. The impact of
the impurities on the temperature of the superfluid transition is
investigated. The impact is found weak at the impurity concentration
smaller than the density of the pairs. It is shown that in the
rarefied system  the critical temperature $T_c\approx 100$ K can be
reached.
\end{abstract}
\maketitle

\section{Introduction}

Recently a series of attempts is made to reveal experimentally the
superfluidity of electron-hole pairs in bilayer systems, where one
layer is of the electron-type conductivity and the other one is the
hole-type one. This phenomenon had been predicted in \cite{1,2,3},
but it was not obtained a fast experimental confirmation.

Discovery of the quantum Hall effect and development of technology
of semiconductor heterostructures allowed to create so called double
quantum wells, i. e. structures with two parallel conducting layers.
Usually in such structures both layers possess  conductivity of the
same type. If a double quantum well is placed in a strong magnetic
field directed normally to the conductive layers, electron-hole
pairing can also take place in it. To achieve this, filling factors
of the Landau levels $\nu_1$ and $\nu_2$ in the layers 1 and 2 must
satisfy the condition $\nu_1+\nu_2=1$. The role of holes is played
by unoccupied states in the zero Landau level and at these filling
factors the concentration of electrons in one layer is equal to the
concentration of holes in the other layer.

Electron-hole pairing in bilayer quantum Hall systems has been predicted in \cite{4,5,6,7}.
The effect has obtained rather convincing experimental confirmation.
Observation of vanishing Hall resistance and a sharp increase of longitudinal conductivity
 at low temperature under a flow of equal by magnitude and oppositely directed currents in the layers
 \cite{8,9,10} can be considered as a direct evidence of the pairing.
Another confirmation is an observable peak in differential
interlayer conductivity at zero potential difference \cite{11,12}.
Its appearance indicates a Josephson nature of interlayer tunneling.
Furthermore, a perfect interlayer drag occurs \cite{13}, i. e. a
process in which currents in the drag and the drive layers are equal
by magnitude. This property naturally follows from the assumption
that electronic transport in such systems is caused by motion of
electron-hole pairs.

The problem of realization of electron-hole pairing without a
magnetic field and superfluidity of gas of such pairs in bilayer
system remains open at present. Quite recently a suggestion has been
made to use graphene layers as components of bilayer systems
\cite{14,15,16,17,18}. Further investigations revealed that a
serious problem in this case is screening of Coulomb interaction
between the electron and the hole \cite{19,20}. The screening effect
is most dangerous in the regime of weak coupling, when the size of
electron-hole pairs significantly exceeds the average interparticle
distance (BCS regime). Screening leads to significant decrease of
the coupling constant, therefore the BCS transition temperature
becomes exponentially small, and presence of even a minor
concentration of impurities, whose influence in this system is
similar to the influence of paramagnetic impurities in ordinary
superconductors \cite{21,22}, leads to almost complete suppression
of electron-hole superfluidity.

A peculiarity of graphene systems that have a Dirac spectrum of
carriers is impossibility of forming local electron-hole pairs in
them. On the contrary, in bilayer systems with a parabolic
dispersion law these pairs can appear. At decreasing the carrier
density there occurs a transition from the BCS regime to the
Bose-Einstein condensation (BEC) regime. In the BEC regime the pairs
are strongly coupled, the pair size is less or much less than the
average distance between the pairs and the screening is suppressed.
Therefore one can expect that the BEC regime will turn out to be
more promising to achieve high superfluid transition temperatures.
It should be noted that in a bilayer graphene system suppression of
screening can also take place. In the case of systems with Dirac
spectrum  a sufficiently high interaction constant $\alpha=e^2/
\hbar v_F\varepsilon$  is necessary for this suppression
\cite{23,24}. Here $v_F$ is the Fermi velocity in graphene and
$\varepsilon$ is the dielectric constant of the matrix or the
effective dielectric constant $\varepsilon_{eff}=(\varepsilon+1)/2$
for a system on a dielectric substrate. The critical value of
$\alpha$, according to various estimates \cite{23,24}, lies in the
range $\alpha_c=1.5 \div  3$. Since for graphene in vacuum
$\alpha=e^2/ \hbar v_F=2.2$, a graphene system on a standard
dielectric substrate with $\varepsilon=4$ obviously does not satisfy
the condition $\alpha>\alpha_c$. This fact probably explains the
negative results of the experiment \cite{25}, where an attempt was
made to find anomalous interlayer drag \cite{26} in a bilayer
graphene system under decrease in temperature.

Anomalous drag has been experimentally observed in electron-hole systems created in
AlGaAs heterostructures \cite{27,28}, and also in hybrid graphene -- AlGaAs structures \cite{29}.
The parameters of the systems used in these experiments correspond to the case of local pairing.
These investigations showed that the drag effect manifests itself stronger with decreasing the carrier density and weaker with increasing the interlayer barrier width.
The first dependence correlates with the fact that in systems with parabolic dispersion law at low carrier density the screening is suppressed and, therefore, the superfluid transition temperature increases.
The second one may indicate destruction of the superfluid state with increasing the interlayer distance.

It is known that superfluidity of pairs in bilayer quantum Hall systems is destroyed with increasing the interlayer distance.
Analysis of the collective excitation spectrum of the system, more precisely, its dependence on the interlayer distance $d$, shows that increasing $d$ leads to appearance of a minimum in the spectrum at finite wave vectors \cite{4,30}.
At $d=d_c$ the dispersion curve touches the abscissa axis.
At zero imbalance of filling factors theory gives the critical distance $d_c\approx 1.2 \ell_H$, where $\ell_H$ is the magnetic length.
At $d>d_c$ the collective mode frequency in a certain range of wave vectors becomes imaginary, that corresponds to instability of the state with pairing.
The magnitude of $d_c$ grows with increasing the imbalance \cite{18}.
Increasing the imbalance leads to decreasing the superfluid density $n_s$, i. e. the increase of $d_c$ can be linked to the decrease of $n_s$.
The presence of a critical distance in the quantum Hall systems and its growth with increasing the imbalance are confirmed experimentally \cite{8,10,31}, although complete quantitative coincidence with theory and experiment is not achieved (the experimental value of $d_c$ is approximately 1.5 times greater than the theoretical one).

The conditions of pairing in bilayer electron-hole systems (without
magnetic field) with low carrier density have been analyzed, in
particular, in \cite{32}. The energy of the system with coupling has
been calculated as a function of pair density, taking into account
exchange, direct Coulomb and Van der Waals interactions. The main
conclusion of \cite{32} is the prediction of a gas-liquid transition
with decreasing the interlayer distance. The gas-liquid transition
means that the gas of pairs becomes unstable to formation of drops
whose density is fixed and independent of the average carrier
density in the system. According to \cite{32}, the drops form only
in sufficiently rarefied systems, furthermore, the distance between
the layers must not exceed a certain limit ($d_c\approx 0.5 a_0$,
where $a_0$ is the effective Bohr radius of the pair). At larger $d$
the gas-liquid transition does not occur. If the average density is
greater than the equilibrium density of the drops ($n_c\approx 0.02
a_0^{-2}$), the gas-liquid transition does not occur at any $d$. The
prediction of the gas-liquid transition correlates with conclusions
of \cite{33,34,35,36}, where  formation of biexcitons in bilayer
systems was discussed. A biexciton consists of two coupled
electron-hole pairs. As it is shown in \cite{34,35,36}, formation of
biexcitons leads to decreasing of the energy of two pairs if the
interlayer distance is less than the critical one. The critical
distance depends on the ratio of electron $m_e$ and hole $m_h$
masses ($d_{c1}\approx 0.9 a_0$ for $m_e\ll m_h$, and $d_{c1}\approx
0.4 a_0$ for $m_e= m_h$). In \cite{33,34,35,36} the question about
coalescence of excitons and biexcitons into drops with large
quantity of excitons was not analyzed. The critical distance
obtained in \cite{32}, also in \cite{34,35,36}, is the lower
critical distance. Collective excitations were not studied in these
articles and any upper limitation on the interlayer distance has not
been obtained.

When studying the electron-hole pairing in bilayer systems the main
interest is the conditions when the gas of pairs is superfluid.
Keldysh  \cite{37} proposed to use a formalism of coherent states to
describe the superfluid state of excitons in the low density limit.
The motivation for using this formalism is the following. In the
theory of Bose gas the superfluid state is described by the order
parameter
\begin{equation}\label{1}
    \Psi(\mathbf{R})=\langle \Phi_0 | \hat{\Psi}(\mathbf{R})|\Phi_0 \rangle,
\end{equation}
where $|\Phi_0 \rangle$ is the ground state wave function of the many-particle system and $\hat{\Psi}(\mathbf{R})$ is the operator of annihilation of a boson at the point $\mathbf{R}$.
The equality (\ref{1}) is obviously satisfied if $|\Phi_0 \rangle$  is an eigenfunction of the operator $\hat{\Psi}(\mathbf{R})$, i. e.
\begin{equation}\label{2}
\hat{\Psi}(\mathbf{R})|\Phi_0 \rangle= \Psi(\mathbf{R})|\Phi_0 \rangle.
\end{equation}
Equation (\ref{2}) can be easily solved.
For that we use the expansion
\begin{equation}\label{3}
\hat{\Psi}(\mathbf{R})=\frac{1}{\sqrt{V}}\sum_\mathbf{k}\hat{a}_\mathbf{k} e^{i\mathbf{k}\mathbf{R}},
\end{equation}
where $\hat{a}_\mathbf{k}$  is the annihilation operator of bosons in the state with wave vector $\mathbf{k}$, and $V$ is the volume of the system.
Similarly we can write down the order parameter
\begin{equation}\label{4}
\Psi(\mathbf{R})=\frac{1}{\sqrt{V}}\sum_\mathbf{k}\alpha_\mathbf{k} e^{i\mathbf{k}\mathbf{R}}.
\end{equation}
As the result, we arrive at necessity to find the eigenfunctions of the annihilation operator $\hat{a}_\mathbf{k}$
\begin{equation}\label{5}
\hat{a}_\mathbf{k}|\alpha_\mathbf{k}\rangle=\alpha_\mathbf{k}|\alpha_\mathbf{k}\rangle.
\end{equation}
These eigenfunctions are well-known \cite{38} and have the form
\begin{equation}
\label{6} |\alpha_\mathbf{k}\rangle=\exp
\left(-\frac{1}{2}|\alpha_\mathbf{k}|^2\right)
\sum_{n=0}^\infty\frac{\alpha^n_\mathbf{k}}{\sqrt{n!}}|n_\mathbf{k}\rangle
    =\exp\left(\alpha_\mathbf{k} \hat{a}^+_\mathbf{k}
    -H.c.\right)|0\rangle,
\end{equation}
where $|n_\mathbf{k}\rangle$ are Fock states.
Functions $|\alpha_\mathbf{k}\rangle$ are called coherent states.
The expressions obtained allow to find easily that
\begin{equation}\label{7}
|\Phi_0
\rangle=\prod_{\mathbf{k}}\exp\left(\alpha_\mathbf{k}\hat{a}^+_\mathbf{k}
-{H.c.}\right)|0\rangle=\exp
\left(\int\Psi(\mathbf{R})\hat{\Psi}^+(\mathbf{R})
d\mathbf{R}-{H.c}\right)|0\rangle\equiv {{D}_B}|0\rangle.
\end{equation}
A natural generalization of expression (\ref{7}) for a system formed by electron-hole pairs is the wave function proposed by Keldysh \cite{37},
\begin{equation}\label{8}
    |\Phi_0\rangle={{D}_F}|0\rangle,
\end{equation}
where $|0\rangle$ is the vacuum state where electrons and holes are absent, and the operator $D_F$ is determined by expression
\begin{equation}\label{9}
    {{D}_F}=\exp\left[\sum_{\sigma,\sigma'}\int d {\bf
    r}_1 d{\bf
    r}_2 \Phi_{\sigma\sigma'}({\bf
    r}_1, {\bf
    r}_2)e^{-i \mu t/\hbar}\psi^{(e)+}_{\sigma}({\bf
    r}_1) \psi^{(h)+}_{\sigma'}({\bf
    r}_2)-H.c.\right].
\end{equation}
In (\ref{9}) $\psi^{(e)+}_{\sigma}({\bf r})$ and
$\psi^{(h)+}_{\sigma}({\bf r})$ are creation operators of an
electron and a hole in corresponding layers, $\sigma$ is the spin
index, $\mu$ is the chemical potential, and
$\Phi_{\sigma\sigma'}({\bf r}_1, {\bf r}_2)$ has the meaning of the
wave function of the pairs.

The formalism of the coherent state has been used in \cite{39a,39}
to describe electron-hole pairing in bilayer quantum Hall systems.
In articles \cite{40,41,42} polarization phenomena in 3D superfluid
gas of electron-hole pairs (without spatial separation of electrons
and holes) have been analyzed using an approach based on Keldysh's
function. In \cite{43} the Keldysh's approach has been applied to
description of the superfluid state of a rarefied gas formed by
alkali metal atoms.

In this article the formalism of coherent states is used to find the
spectrum of collective excitations in a superfluid gas of
electron-hole pairs in the absence of magnetic fields and to analyze
stability of the superfluid state. It is shown that under increase
of the distance $d$ between the layers a minimum appears in the
excitation spectrum, and at a certain critical value $d_ñ$ the
dispersion curve touches the abscissa axis and the excitation energy
becomes imaginary. In it found that the critical distance $d_c$
increases under decrease in density of carriers. In the last
section, within the same formalism, we consider the influence of
impurities on the superfluid transition temperature.

\section{Energy of the electron-hole coherent state}

The Hamiltonian of a bilayer system consisting of electron and hole layers has the form
\begin{eqnarray}\label{10}
    H=-\sum_{\alpha=e,h,\sigma=\uparrow,\downarrow}\int d {\bf r}\frac{\hbar^2}{2
    m_\alpha}\psi^{(\alpha)+}_{\sigma}({\bf r})\nabla^2
    \psi^{(\alpha)}_{\sigma}({\bf r})\cr
    +\frac{1}{2}\sum_{\alpha,\beta=e,h,\sigma,\sigma'=\uparrow,\downarrow}
    \int d {\bf
    r}d{\bf
    r'}\psi^{(\alpha)+}_{\sigma}({\bf r})
    \psi^{(\beta))+}_{\sigma'}({\bf r'})
    V_{\alpha\beta}(|{\bf r}-{\bf r'}|)
    \psi^{(\beta)}_{\sigma'}({\bf r'})\psi^{(\alpha)}_{\sigma}({\bf
    r}),
\end{eqnarray}
where $m_e$ and $m_h$ are effective masses of an electron and a hole,
$V_{\alpha\beta}(r)$ is the interaction energy between the carriers and $\mathbf{r}$
is a two-dimensional radius vector.
Assume that the bilayer system is placed in a homogeneous dielectric
matrix with dielectric constant $\varepsilon$ coincident with
 the dielectric constant of the interlayer between the electron and hole layers.
In this case $V_{ee}(r)=V_{hh}(r)=e^2/\varepsilon r$ and
$V_{eh}(r)=-e^2/\varepsilon \sqrt{r^2+d^2}$.

The wave function of pairs in (\ref{9}) has a matrix structure.
We consider now singlet pairing.
This corresponds to the matrix $\Phi_{\sigma\sigma'}({\bf r}_1, {\bf r}_2)$,
where only components non-diagonal by spin indexes are nonzero, i. e. pairing is described by two scalar wave functions $\Phi_{\uparrow\downarrow}({\bf r}_1, {\bf r}_2)$ and $\Phi_{\downarrow\uparrow}({\bf r}_1, {\bf r}_2)$ (in following, for short, we use the symbol $\Phi_\sigma\equiv\Phi_{\sigma,-\sigma}$).
In the general case these functions are different.
If both these functions are nonzero, the gas of pairs is two-component.
The first component corresponds to an electron with spin (spin projection)
 $+1/2$ coupled to a hole with spin $-1/2$, the second one -- to an electron with spin $-1/2$ coupled to a hole with spin $+1/2$.
In this case the operator $D_F$ has the form
\begin{equation}\label{11}
    {D_F}=\exp\left[\sum_{\sigma}\int d {\bf
    r}_1 d{\bf
    r}_2 \Phi_{\sigma}({\bf
    r}_1, {\bf
    r}_2)e^{-i \mu_\sigma t/\hbar}\psi^{(e)+}_{\sigma}({\bf
    r}_1) \psi^{(h)+}_{-\sigma}({\bf
    r}_2)-H.c.\right],
\end{equation}
where values $\mu_\sigma$ are chemical potentials of the components.

It is known that a two-component Bose gas is unstable relative to
separation into the components if the square of the interaction
constant between pairs of different types exceeds the product of
interaction constants between pairs of the same types \cite{44}.
Depending on the relation between the interaction constants, the
condensate will be either a homogeneous mixture of pairs of two
types or a biphasic system with only one component present in each
phase.

Functions $\Phi_\sigma$ can be found from the condition of minimum of the functional
\begin{equation}\label{12}
    F=E-\sum_\sigma\mu_\sigma N_\sigma,
\end{equation}
where $E$ and $N_\sigma$ are the energy of the system and the number of pairs in the $\sigma$ component in the state ({\ref{8}).
Their values are determined by expressions
\begin{equation}\label{13}
E= \langle 0|\tilde{H}|0\rangle
\end{equation}
and
\begin{equation}\label{14}
N_\sigma= \langle 0| \tilde{ N}_\sigma|0\rangle,
\end{equation}
where $\tilde{H}$ has the form coinciding with the initial Hamiltonian (\ref{10}) with operators $\psi^{(\alpha)}_{\sigma}({\bf r})$ in it replaced by
\begin{equation}\label{15}
        \tilde{\psi}^{(e,h)}_{\sigma}({\bf r})=D_F^+\psi^{(e,h)}_{\sigma}({\bf
        r}){D_F},
\end{equation}
and the pair number operator has the form
\begin{equation}\label{16}
   \tilde{ N}_\sigma=\int d {\bf r}
\tilde{\psi}^{(e)+}_{\sigma}({\bf
r})\tilde{\psi}^{(e)}_{\sigma}({\bf r})=\int d {\bf r}
\tilde{\psi}^{(h)+}_{-\sigma}({\bf
r})\tilde{\psi}^{(h)}_{-\sigma}({\bf r}).
\end{equation}
Taking into account the explicit form of $D_F$, we can express the operators $\tilde{\psi}^{(\alpha)}_{\sigma}$ in terms of creation and annihilation of electrons and holes in the following way \cite{37}
\begin{eqnarray}
\label{17}
  \tilde{\psi}^{(e)}_{\sigma}({\bf r}) &=& \int d {\bf r'}[C^{(e)}_{\sigma}({\bf r},{\bf r'})
  {\psi}^{(e)}_{\sigma}({\bf r'})
  +e^{-i\mu_\sigma t/\hbar}S_\sigma({\bf r},{\bf r'}){\psi}^{(h)+}_{-\sigma}({\bf r'})], \cr
  \tilde{\psi}^{(h)}_{-\sigma}({\bf r}) &=&
  \int d {\bf r'}[C^{(h)}_{\sigma}({\bf r'},{\bf r}){\psi}^{(h)}_{-\sigma}({\bf r'})
  -e^{-i\mu_\sigma t/\hbar}S_\sigma({\bf r'},{\bf r}){\psi}^{(e)+}_{\sigma}({\bf
  r'})],
\end{eqnarray}
where
\begin{eqnarray}\label{18}
 C^{(e)}_{\sigma}({\bf r},{\bf r'}) &=& \delta(\mathbf{r}-\mathbf{r}')+\sum_{n=1}^\infty \frac{(-1)^n}{(2n)!}
 (\Phi_\sigma\cdot\Phi^+_\sigma)^n,\cr
 C^{(h)}_{\sigma}({\bf r},{\bf r'})&=& \delta(\mathbf{r}-\mathbf{r}')+\sum_{n=1}^\infty \frac{(-1)^n}{(2n)!}
 (\Phi^+_\sigma\cdot\Phi_\sigma)^n,\cr
  S_\sigma({\bf r},{\bf r'}) &=& \sum_{n=0}^\infty\frac{(-1)^n}{(2n+1)!}\Phi_\sigma \cdot
  (\Phi^+_\sigma\cdot\Phi_\sigma)^n.
\end{eqnarray}
In (\ref{18}) we used a notation $\Phi^+_\sigma({\bf r}_1,{\bf r}_2)=\Phi^*_\sigma({\bf r}_2,{\bf r}_1)$ and the product sign means a convolution.

The functional (\ref{12}) is an infinite series containing convolutions of $\Phi_\sigma({\bf r},{\bf r}')$ of even orders.
In the low density limit we can limit ourselves to the terms up to the fourth order inclusive.
With only the second order terms taken into account, the condition that the variation of the functional $F$ equals to zero gives a Schr\"odinger equation for a separate electron-hole pair
\begin{equation}\label{19}
  \left[-\frac{\hbar^2}{2
m_e}\nabla^2_1-\frac{\hbar^2}{2 m_h}\nabla^2_2 +V_{eh}(|{\bf
r}_1-{\bf r}_2|)-\mu_\sigma\right]\Phi_\sigma({\bf r}_1,{\bf
r}_2)=0.
\end{equation}
The solution of this equation can be written  in the form
\begin{equation}\label{20}
\Phi_\sigma({\bf r}_1,{\bf r}_2)=\Psi_\sigma({\bf R})\phi({\bf r}),
\end{equation}
where ${\bf R}$ is the center of mass coordinate, ${\bf r}$ is the relative coordinate, $\Psi_\sigma({\bf R})$ is the wave function of the pair moving as a whole, and $\phi({\bf r})$ is the function describing the bound electron-hole state.
The function $\phi({\bf r})$ satisfies an equation
\begin{eqnarray}
\label{21}
 -\frac{\hbar^2}{2 m}\nabla^2_r \phi({\bf r})+ V_{eh}(r) \phi({\bf r})&=&
\mu_\sigma\phi({\bf r}),
\end{eqnarray}
where $m=m_e m_h/(m_e+m_h)$ is the reduced mass.
Equation (\ref{21}) is the Schr\"odinger equation for a particle in an isotropic two-dimensional potential.
The energy minimum is achieved for the ground state wave function $\phi_0({\bf r})$.
In this approximation the chemical potentials $\mu_\uparrow$ and $\mu_\downarrow$ coincide
and are equal to the ground state energy $E_0$.
Normalization of functions $\Phi_\sigma({\bf r},{\bf r}')$ is given by the condition
\begin{equation}\label{22}
 \sum_\sigma\int d {\bf R} d{\bf r} |\Psi_\sigma({\bf R})|^2 |\phi_0({\bf r})|^2 =N,
\end{equation}
where $N$ is the total number of pairs.

Equation (\ref{22}) leaves an arbitrariness in choosing the normalization of functions $\Phi_\sigma({\bf R})$ è $\phi_0({\bf r})$.
For definiteness, we will assume $\int d{\bf r} |\phi_0({\bf r})|^2 =1$ and $\sum_\sigma\int d{\bf R} |\Psi_\sigma({\bf R})|^2 =N$.
In the ground state $\Psi_\sigma({\bf R})=\Psi_{0\sigma}=\sqrt{n_\sigma}$, where $n_\sigma$ is the density of pairs of type $\sigma$.

Let us now consider the fourth order terms in the functional
(\ref{12}). We will seek the functions $\Phi_\sigma$ corresponding
to the minimum of the functional (\ref{12}) in the form
$\Phi_\sigma({\bf r}_1,{\bf r}_2)=\Psi_\sigma({\bf
R}_{12})\phi_0({\bf r}_{12})$. At the same time we neglect the
correction to the function $\phi_0$ caused by interaction between
the pairs. In this approximation
\begin{eqnarray}\label{23}
F=\sum_\sigma\int d {\bf R} \Psi^*_\sigma({\bf
R})\left[-\frac{\hbar^2}{2 M}\nabla^2_R
-\tilde{\mu}_\sigma\right]\Psi_\sigma({\bf R})\cr +
\frac{1}{2}\sum_{\sigma,\sigma'}\int d{\bf r}_1 d {\bf r}_2 d {\bf
r}_3 d {\bf r}_4 A\left( {\bf r}_1,{\bf r}_2,{\bf r}_3,{\bf r}_4
\right) |\Psi_\sigma( {\bf R}_{12})|^2 |\Psi_{\sigma'}( {\bf
R}_{34})|^2+\cr \frac{1}{2}\sum_\sigma\int d {\bf r}_1 d {\bf r}_2 d
{\bf r}_3 d {\bf r}_4 B\left({\bf r}_1,{\bf r}_2,{\bf r}_3,{\bf r}_4
\right)\Psi^*_\sigma( {\bf R}_{12})\Psi_\sigma( {\bf
R}_{32})\Psi^*_\sigma( {\bf R}_{34})\Psi_\sigma( {\bf R}_{14}),
\end{eqnarray}
where functions $A[{\bf r}_i]$ and $B[{\bf r}_i]$ are expressed in terms of the Coulomb interaction potential and the wave function of the bound electron-hole state
\begin{equation}\label{24}
 A[{\bf r}_i]=V_d({\bf r}_1,{\bf r}_2,{\bf r}_3,{\bf r}_4 )|\phi_0( {\bf
r}_{12})|^2 |\phi_0( {\bf r}_{34})|^2,
\end{equation}
\begin{equation}\label{25}
 B[{\bf r}_i]=-V_{ex}({\bf r}_1,{\bf r}_2,{\bf r}_3,{\bf r}_4 )\phi_0^*( {\bf r}_{12})\phi_0( {\bf r}_{32})\phi_0^*( {\bf
r}_{34})\phi_0( {\bf r}_{14}).
\end{equation}
In (\ref{24}), (\ref{25}) we use the notations
\begin{equation}\label{26}
  V_d({\bf r}_1,{\bf r}_2,{\bf r}_3,{\bf r}_4
)=V_{ee}({ r}_{13})+V_{hh}({ r}_{24})+V_{eh}({
r}_{14})+V_{eh}({r}_{23})
\end{equation}
and
\begin{equation}\label{27}
  V_{ex}({\bf r}_1,{\bf r}_2,{\bf r}_3,{\bf r}_4
)=V_{ee}({ r}_{13})+V_{hh}({ r}_{24})+\frac{1}{2}\left[(V_{eh}({
r}_{12})+V_{eh}({ r}_{34})+V_{eh}({ r}_{14})+V_{eh}({
r}_{23})\right].
\end{equation}
The value $\tilde{\mu}_\sigma$ in (\ref{23}) is the shift of the chemical potential caused by interaction between the pairs, $\tilde{\mu}_\sigma=\mu_\sigma-E_0$.

The extremum condition of the functional (\ref{23}) leads to an equation
\begin{eqnarray}\label{28}
\tilde{\mu}_\sigma \Psi_\sigma({\bf R}_{12})= -\frac{\hbar^2}{2
M}\nabla^2_{R_{12}}\Psi_\sigma({\bf R}_{12})\cr +\int d {\bf r}_{12}
d {\bf r}_3 d {\bf r}_4 \left[A[{\bf r}_i] \Psi_\sigma({\bf
R}_{12})(\sum_{\sigma'}|\Psi_{\sigma'}( {\bf R}_{34})|^2) + B[{\bf
r}_i]\Psi_\sigma( {\bf R}_{32})\Psi^*_\sigma( {\bf
R}_{34})\Psi_\sigma( {\bf R}_{14})\right],
\end{eqnarray}
where $M=m_e+m_h$ is the pair mass.
Equation (\ref{28}) is satisfied by a function $\Psi_\sigma({\bf R})=\sqrt{n_\sigma}$ with an appropriate choice of $\tilde{\mu}_\sigma$.
Substituting $\Psi_\sigma$ and $\tilde{\mu}_\sigma$ found into (\ref{23}), we obtain
\begin{equation}\label{29}
F= -\frac{S}{2}\left[\gamma_{11}(n^2_\uparrow +
n^2_\downarrow)+2\gamma_{12}n_\uparrow n_\downarrow\right],
\end{equation}
where $S$ is the area of the system,
 $\gamma_{11}=\gamma_d+\gamma_{ex}$ and $\gamma_{12}=\gamma_{d}$ are
 interaction constants between the pairs of the same type and different types correspondingly.
These  constants contain the contributions of direct ($\gamma_d$)
and exchange ($\gamma_{ex}$) Coulomb interactions,
\begin{equation}\label{30}
   \gamma_d=\frac{4\pi e^2 d}{\varepsilon},
\end{equation}
\begin{equation}\label{31}
 \gamma_{ex}= -\frac{4\pi e^2}{\varepsilon}
\int\frac{d^2 p}{(2\pi)^2}\frac{d^2
 q}{(2\pi)^2}\frac{1}{p}|\phi_{\mathbf{q}}|^2\Bigg[
|\phi_{\mathbf{q} +\mathbf{p}}|^2-\frac{e^{- p
d}}{2}\left(\phi^*_{\mathbf{q}+\mathbf{p}}\phi_{\mathbf{q}}+
\phi^*_{\mathbf{q}}\phi_{\mathbf{q}+\mathbf{p}}\right)\Bigg],
\end{equation}
where $\phi_{\mathbf{q}}=\int d\mathbf{r} e^{i \mathbf{q}\mathbf{r}}\phi_0(\mathbf{r})$ is the Fourier component of the bound state wave function.

For fixed total density $n=n_\uparrow+n_\downarrow$ we find that at
$\gamma_{ex}>0$ the minimum of the functional (\ref{29}) corresponds
to a homogeneous mixture of components
$n_\uparrow=n_\downarrow=n/2$, and at $\gamma_{ex}<0$ the minimum is
reached if
 $n_\uparrow=n$, $n_\downarrow=0$ or $n_\uparrow=0$, $n_\downarrow=n$.
The situation, when in the whole system the density of only one
component is nonzero, corresponds to complete spin polarization of
the electron and hole layers. To minimize the magnetic energy,
appearance of a domain structure with regions possessing opposite
signs of polarization is preferable. In each region the density of
only one component will be nonzero, but averaged densities of both
components in the whole system will be equal. Under conservation of
the average densities of the spin components the formation of the
domain structure can be also interpreted as  spatial separation of
the components.

It can be shown \cite{41} that at $V_{ee}(r)=V_{hh}(r)=-V_{eh}(r)$
(that corresponds to $d=0$) the value of $\gamma_{ex}$ is positive.
However, at $d$ exceeding a certain critical value, the coefficient
$\gamma_{ex}$ will change its sign. The reason is that at large $d$
the contribution of the first term into the integral in (\ref{31})
becomes dominating.

 At $d=0$ equation (\ref{21}) is a
two-dimensional Schr\"odinger equation for a charged particle in a
Coulomb field. The ground state solution of this equation  has the
form
\begin{equation}\label{32}
    \phi_0(r)=\frac{1}{a_0}\sqrt{\frac{8}{\pi}} e^{-2 r/a_0},
\end{equation}
where $a_0=\hbar^2\varepsilon/m e^2$ is the effective Bohr radius of the pair.
The Fourier component of the function (\ref{32}) equals to
\begin{equation}\label{33}
\phi_\mathbf{q}=\frac{\sqrt{2\pi}a_0}{\left(1+\frac{q^2 a_0^2}{4}\right)^{3/2}}.
\end{equation}
Substitution of the function (\ref{33}) into (\ref{31}) at $d=0$
yields $\gamma_{ex}= +{3.03 e^2 a_0}/{\varepsilon}$. At $d\gtrsim
a_0$ the potential $V_{er}(r)$ can be replaced with a harmonic one
$V_{eh}(r)\approx -e^2/\varepsilon d+e^2r^2/2 \varepsilon d^3$ and
the ground state wave function can be written in the form

\begin{equation}\label{34}
    \phi_0({\bf r})=\frac{1}{\sqrt{\pi}r_0}e^{-\frac{r^2}{2 r_0^2}}.
\end{equation}
Here $r_0=\sqrt[4]{a_0d^3}$ is a length parameter which should be interpreted as a characteristic pair size.
The Fourier component of (\ref{34}) is
\begin{equation}\label{35}
\phi_\mathbf{q}=\sqrt{4\pi}r_0\exp\left(-\frac{q^2 r_0^2}{2}\right).
\end{equation}
Taking $d=a_0$ and substituting (\ref{35}) into (\ref{31}), we
obtain $\gamma_{ex}= -{6.96 e^2 a_0}/{\varepsilon}$. This shows that
the sign change of the constant $\gamma_{ex}$ occurs at $d<a_0$. To
estimate $d$ at which the sign change occurs, we can substitute into
(\ref{31}) the function (\ref{33}) assuming that at small $d$ it
does not significantly differ from the exact function. The
dependence obtained, represented in Fig. \ref{f1}, shows that the
sign change of $\gamma_{ex}$ occurs at a sufficiently small
interlayer distance $d\approx 0.2 a_0$. For greater $d$ we predict
spatial separation of superfluid components.

\begin{figure}
\begin{center}
\includegraphics[width=8cm]{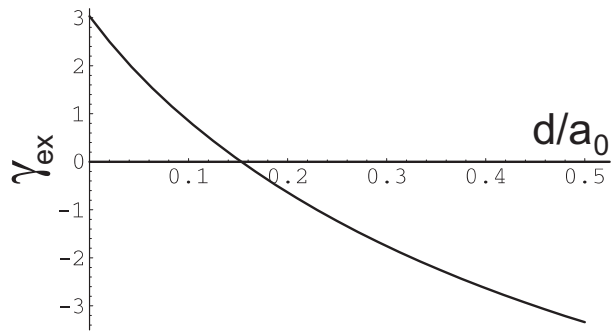}
\end{center}
\caption{Dependence of the exchange part of the interaction constant (in the units $e^2 a_0/\varepsilon$) on the interlayer distance.}
\label{f1}
\end{figure}

We note that in recent paper \cite{md} it was considered the
possibility of electron-hole pairing in a double layer system formed
by two-dimensional transition metal dichalcogenides (TMD) that are
separated by hexagonal boron nitride. The results of that paper
obtained from the analysis of the equations for the order parameters
of pairing and for the chemical potential are in correlation with
our results. Two-dimensional crystals of TMD have the honeycomb
lattice similar to graphene one. The minima of the conductivity band
and the maxima of the valence band are localed in the $K$ and $K'$
points of the Brillouin zone. A strong spin splitting occurs the
valence band. In a result, the pairs of two species can emerge. The
species distinguish by the spin and valley indexes. It was shown in
\cite{md} that at $d< 0.25 a_0$ the energy minimum corresponds to
the two-component electron-hole pair state, while at $d> 0.25 a_0$
the spin-polarized state one component state is realized.

\section{Spectrum of collective excitations and critical interlayer distance}

A homogeneous two-component Bose condensate has two collective modes
whose dispersion laws in the long wavelength limit are acoustic:
$\omega_{\pm}(q)=s_{\pm} q$. If the component densities are
identical and equal to $n_1=n_2=n/2$, sound velocities are
determined by an expression $s_{\pm}=\sqrt{(\gamma_{11}\pm
\gamma_{12})n/{2M}}$ (see e. g. \cite{45}). For a two-component
condensate of pairs
\begin{equation}\label{36}
  s_+=\sqrt{\frac{(2\gamma_{d}+\gamma_{ex})n}{2M}},\quad
  s_-=\sqrt{\frac{\gamma_{ex}n}{2M}}.
\end{equation}
The quantity $s_-$ is real if $\gamma_{ex}>0$. In the opposite case,
$\gamma_{ex}<0$, a homogeneous two-component phase will be unstable
relative to spatial separation into components. If there is only one
superfluid component in a given region, one collective mode
corresponds to it. The spectrum of this mode at small wave vectors
is acoustic with sound velocity
\begin{equation}\label{37}
  s=\sqrt{\frac{(\gamma_{d}+\gamma_{ex})n}{M}}.
\end{equation}
In the general case the sum $\gamma_{d}+\gamma_{ex}=\gamma_{11}$ can
also be negative. In this situation the layered phase would be
unstable relative to formation of drops of dense phase, however, as
it is shown further, in our model $\gamma_{11}$ remains positive for
all $d$.

Now let us proceed to finding the excitation spectrum at finite wave
vectors. We limit ourselves to the case of separated components. We
use the equation (\ref{28}) modified considering that in a given
region of space there is only one component present. Interaction
between components at large distances is neglected. We consider the
time-dependent function $\Psi$ and replace the left-hand side of
(\ref{28}) with its time derivative (here and below we will omit the
component index). As the result we arrive at an equation
\begin{eqnarray}\label{38}
i \hbar \frac{\partial}{\partial t} \Psi({\bf R}_{12},t)=
-\frac{\hbar^2}{2 M}\nabla^2_{R_{12}}\Psi({\bf R}_{12},t)\cr+\int d
{\bf r}_{12} d {\bf r}_3 d {\bf r}_4 \left[A[{\bf r}_i] \Psi({\bf
R}_{12},t)|\Psi( {\bf R}_{34},t)|^2 + B[{\bf r}_i]\Psi( {\bf
R}_{32},t)\Psi^*( {\bf R}_{34},t)\Psi( {\bf R}_{14},t)\right].
\end{eqnarray}
This equation is a modified variant of the Gross-Pitaevskii equation.

The function $\Psi({\bf R},t)$ can be written as a sum of a homogeneous solution and a small correction which is a monochromatic plane wave,
\begin{equation}\label{39}
\Psi({\bf R},t)=\sqrt{n}+e^{-\frac{i \tilde{\mu}
t}{\hbar}}\left(u_{\mathbf{k}} e^{i({\bf k}\cdot{\bf R}-\omega
t)}+v_{\mathbf{k}}^* e^{-i({\bf k}\cdot{\bf R}-\omega t)}\right).
\end{equation}
The chemical potential in (\ref{39}) is found from (\ref{28}) and equals $\tilde{\mu}=(\gamma_d+\gamma_{ex})n$.
Using (\ref{38}) in the linear approximation in the coefficients $u$ and $v$, we obtain a system of equations for these coefficients
\begin{equation}\label{40}
    \left(
      \begin{array}{cc}
        \epsilon_k
        +(\gamma_d(\mathbf{k})+\gamma_{ex}^{(1)}(\mathbf{k}))n
       & (\gamma_d(\mathbf{k})+\gamma_{ex}^{(2)}(\mathbf{k}))n
        \\
       (\gamma_d(\mathbf{k})+\gamma_{ex}^{(2)}(\mathbf{k}))n&
        \epsilon_k
        +(\gamma_d(\mathbf{k})+\gamma_{ex}^{(1)}(\mathbf{k}))n\\
      \end{array}
    \right)\left(
               \begin{array}{c}
                 u_{\mathbf{k}} \\
                 v_{\mathbf{k}} \\
                               \end{array}
             \right)=\hbar\omega\left(
               \begin{array}{c}
                 u_{\mathbf{k}} \\
                 -v_{\mathbf{k}} \\
               \end{array}
             \right).
\end{equation}
Here  $\epsilon_k=\hbar^2 k^2/2 M$,
\begin{equation}\label{41}
     \gamma_d(\mathbf{k})=\int d {\bf r}_{12} d {\bf r}_3 d {\bf r}_4  A[{\bf
    r}_i] e^{i {\bf k}\cdot({\bf R}_{34}-{\bf R}_{12})},
\end{equation}
\begin{equation}\label{42}
  \gamma_{ex}^{(1)}({\bf k})=\int d {\bf r}_{12} d {\bf r}_3 d
{\bf r}_4 B[{\bf
    r}_i]\left(e^{i {\bf k}\cdot({\bf R}_{32}-{\bf R}_{12})}+e^{i {\bf k}\cdot({\bf R}_{14}-{\bf
    R}_{12})}-1\right)
\end{equation}
and
\begin{equation}\label{43}
  \gamma_{ex}^{(2)}({\bf k})=\int d {\bf r}_{12} d {\bf r}_3 d {\bf r}_4 B[{\bf
    r}_i]e^{i {\bf k}\cdot({\bf R}_{34}-{\bf R}_{12})}.
\end{equation}
Values of $\gamma_d({\bf k})$, $\gamma_{ex}^{(1,2)}({\bf k})$ in the system under consideration depend only on the absolute value of the wave vector.

Equating the determinant of the system (\ref{40}) to zero, we find the collective mode spectrum
\begin{equation}\label{44}
   \hbar \omega(k) =\sqrt{\left(\epsilon_k+
    [\gamma_{ex}^{(1)}(k)-\gamma_{ex}^{(2)}(k)]n\right)
    \left(\epsilon_k+
    [2\gamma_d(k)+\gamma_{ex}^{(1)}(k)+\gamma_{ex}^{(2)}(k)]n\right)}.
\end{equation}

Functions (\ref{41}) -- (\ref{43}) can be expressed in terms of the
Fourier component of the wave function $\phi_0({\bf r})$. In the
general case the corresponding expressions have a rather cumbersome
form. We give them in the Appendix. In the $k\to 0$ limit these
quantities are reduced to the constants introduced earlier,
$\gamma_d(0)=\gamma_d$,
$\gamma_{ex}^{(1)}(0)=\gamma_{ex}^{(2)}(0)=\gamma_{ex}$. Using
expression (\ref{35}) for the function $\phi_0(\mathbf{q})$, that
corresponds to the limit of large $d$, and limiting ourselves to the
case of equal electron and hole masses, we obtain the following
analytical expressions for the functions (\ref{41})-(\ref{43}),
\begin{eqnarray}\label{45}
 \gamma_d(k)=\frac{4\pi e^2}{\varepsilon k}(1-e^{- k
 d})e^{-\frac{k^2 r_0^2}{8}},
\end{eqnarray}
\begin{eqnarray}\label{46}
 \gamma_{ex}^{(1)}(k)=-\frac{4\pi e^2 r_0}{\varepsilon }
\Bigg[\sqrt{\frac{\pi}{2}}\left(e^{-\frac{k^2
r_0^2}{16}}I_0\left(\frac{k^2 r_0^2}{16}\right)+e^{-\frac{k^2
r_0^2}{8}}-1\right)-2 e^{-\frac{k^2 r_0^2}{8}} f\left({d},k/4
\right) +f\left({d},0\right)\Bigg],
\end{eqnarray}
\begin{eqnarray}\label{47}
 \gamma_{ex}^{(2)}(k)=-\frac{4\pi e^2 r_0}{\varepsilon }
\Bigg[\sqrt{\frac{\pi}{2}}e^{-\frac{3 k^2
r_0^2}{16}}I_0\left(\frac{k^2 r_0^2}{16}\right)-e^{-\frac{k^2
r_0^2}{4}}\frac{f\left({d},{k }/{2}\right)+f\left({d},0\right)}{2}
\Bigg].
\end{eqnarray}
In these expressions $I_0(x)$ is the modified Bessel function, and $f(d,k)$ is defined in terms of the integral
\begin{equation}\label{48}
    f(d,k)=\int_0^\infty e^{-\frac{3p^2}{8}-\frac{p d}{r_0}}I_0\left(p k r_0\right) d p.
\end{equation}
At $k=0$ this function can be written using the complementary error function
\begin{equation}\label{49}
 f(d,0)=\sqrt{\frac{2 \pi}{3}}\exp\left(\frac{2 d^2}{3
    r_0^2}\right)\mathrm{erfc}\left(\sqrt{\frac{2}{3}}
    \frac{d}{r_0}\right).
\end{equation}
The analytical expression for the constant $\gamma_{ex}$ calculated using the function (\ref{49}) has the form
\begin{eqnarray}\label{50}
 \gamma_{ex}=-\frac{4\pi e^2 r_0}{\varepsilon }
\Bigg[\sqrt{\frac{\pi}{2}}-\sqrt{\frac{2 \pi}{3}}\exp\left(\frac{2
d^2}{3
    r_0^2}\right)\mathrm{erfc}\left(\sqrt{\frac{2}{3}}
    \frac{d}{r_0}\right) \Bigg].
\end{eqnarray}
This expression is valid also for an arbitrary ratio of electron and hole masses.

Fig. \ref{f2} shows the dependence of the constant $\gamma_{11}=\gamma_d+\gamma_{ex}$ on $d$.
For $d>0.5 a_0$ the value of $\gamma_{ex}$ is obtained from (\ref{50}), and for $d<0.5a_0$ -- from (\ref{31}) using the function (\ref{33}) corresponding to the limit $d\to 0$.
Apparently, the dependences join sufficiently fine.
Positivity of the constant $\gamma_{11}$ at all $d$ means that the approximation used in this article does not predict an instability of the system relative to formation of drops (gas-liquid transition).

\begin{figure}
\begin{center}
\includegraphics[width=8cm]{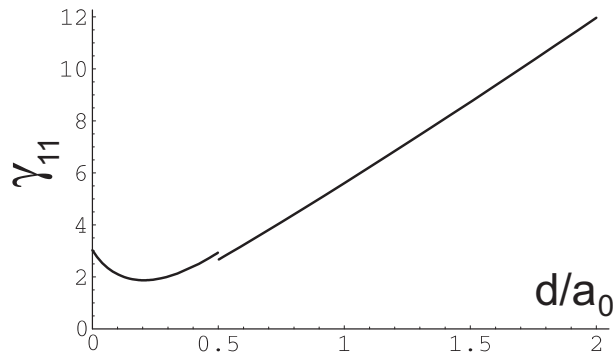}
\end{center}
\caption{Dependence of the interaction constant between the pairs
(in units $e^2 a_0/\varepsilon$) on the interlayer distance. For
$d<0.5 a_0$, $\gamma_{11}$ is computed using (\ref{32}), for $d>0.5
a_0$, using (\ref{34}).}\label{f2}
\end{figure}

Let us now consider the character of change of the collective mode spectrum with variation of density and interlayer distance.
The collective mode spectrum calculated using functions (\ref{45}) -- (\ref{47}) at fixed $d$ and variable density is represented in Fig. \ref{f3}.
Fig. 4 represents the change of the spectrum at fixed density with variation of interlayer distance.
It follows from these dependences that when the density increases, or when the interlayer distance increases, the minimum in the spectrum becomes deeper.
At reaching the critical density $n_c$ or the critical interlayer distance $d_c$ the curve touches the X axis, and after exceeding the critical value the spectrum becomes imaginary.
The latter means that the superfluid state becomes unstable.
The distance $d_c$ depends on $n$, and the density $n_c$ depends on $d$.
The dependence $n_c(d)$ calculated using (\ref{45}) -- (\ref{47}) is shown in Fig. \ref{f5}.
According to this figure, the dependence is a power-law one,
\begin{equation}\label{51}
    n_c a_0^2 \approx C_1 \left(\frac{d}{a_0}\right)^\alpha,
\end{equation}
where the exponent is $\alpha=-2.62$ and the numeric multiplier is $C_1=0.335$.

\begin{figure}
\begin{center}
\includegraphics[width=8cm]{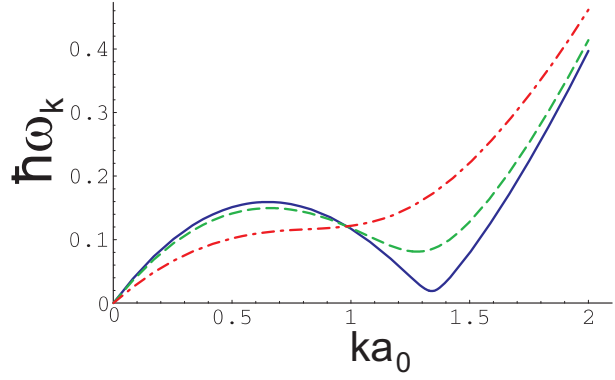}
\end{center}
\caption{Spectrum of excitations in the superfluid gas of electron-hole pairs at $d=1.5 a_0$ for  $na_0^2=0.115$; 0.1; 0.05 (solid, dashed and dash-dot curves correspondingly).
The energy is given in units $\hbar^2/m a_0^2$ (doubled effective Rydberg). }
\label{f3}
\end{figure}

\begin{figure}
\begin{center}
\includegraphics[width=8cm]{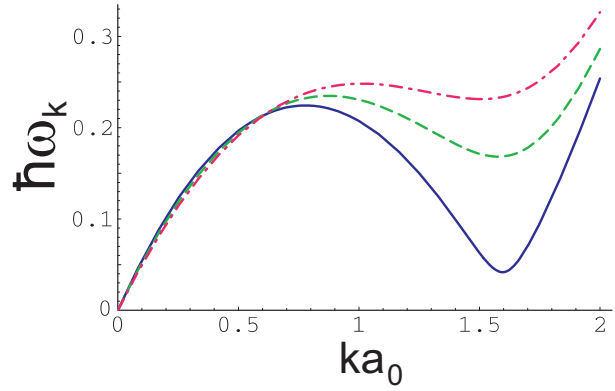}
\end{center}
\caption{Spectrum of excitations in the superfluid gas of electron-hole pairs at $n a_0^2=0.2$ for $d /a_0=1.21$; 1.1; 1.0 (solid, dashed and dash-dot curves correspondingly).
The energy is in the same units as in Fig. \ref{f3}.}
\label{f4}
\end{figure}

\begin{figure}
\begin{center}
\includegraphics[width=8cm]{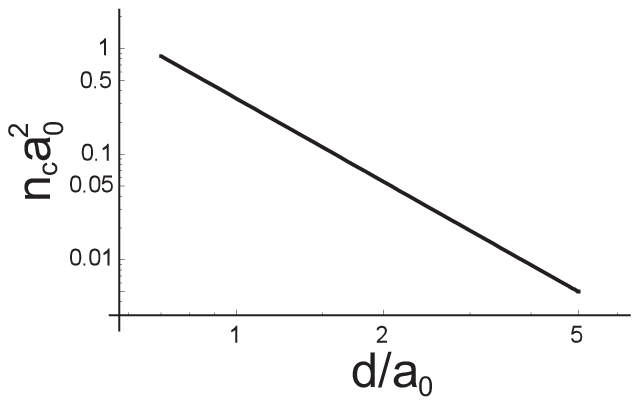}
\end{center}
\caption{Dependence of critical pair density on the interlayer distance in double logarithmic scale.}
\label{f5}
\end{figure}

The low density limit corresponds to pair size lower than the average distance between the pairs.
This means that the formalism used in the article is applicable if the following condition is satisfied:
\begin{equation}\label{52}
    n a_0^2\lesssim \Bigg\{ \begin{array}{cc}
                   \left(\frac{d}{a_0}\right)^{-3/2},&d> a_0; \\
                 1, & d< a_0.
               \end{array}
\end{equation}
In other words, it makes sense to talk about the critical density
(\ref{51}) only if the density satisfies the inequality (\ref{52}).
This takes place if $d \gtrsim 0.7 a_0$. At lower $d$ the expression
(\ref{51}) is inapplicable. One may expect that with increasing the
density a BEC-BCS crossover may occur and not a phase transition
with formation of a density wave.

Instability connected with appearance of a soft mode can be interpreted as instability related to formation of a Wigner crystal.
It is interesting to compare the condition (\ref{51}) with the condition of formation of such a crystal that can be obtained from semiclassical considerations.
The density corresponding to the transition into the crystal phase is by order of magnitude equal to the density at which the average kinetic energy is lower than the dipole-dipole interaction energy., i. e. $\hbar^2 n/M\lesssim e^2 d^2 n^{3/2}/\varepsilon$.
This gives
\begin{equation}\label{53}
  n_c a_0^2 \approx \frac{m}{M}\left(\frac{d}{a_0}\right)^{-4}.
\end{equation}
Comparing (\ref{53}) to (\ref{51}), we arrive at a conclusion that the semiclassical approach underestimates the critical density.

\section{Critical temperature and influence of impurities on it}

Now let us estimate the superfluid transition temperature in the system under study.
For a two-dimensional system this transition is a Berezinskii-Kosterlitz-Thouless transition and its temperature is determined by the equation
\begin{equation}\label{54}
    T_c=\frac{\pi}{2}\frac{\hbar^2 n_s(T_c)}{M},
\end{equation}
where $n_s(T)$ is the superfluid density at temperature $T$.
The superfluid density can be found as a difference between the total pair density $n$ and the normal component density
\begin{equation}\label{55}
   n_s(T)=n-\frac{1}{2}\frac{\hbar^2}{M T}\int \frac{d{\bf k}}{(2\pi)^2}
   k^2 N_B(\hbar\omega_k)[1+N_B(\hbar\omega_k)],
\end{equation}
where $N_B(E)=(e^{E/T}-1)^{-1}$ is the Bose distribution function.
The main contribution into the integral in (\ref{55}) is made by
long wavelength excitations. Approximating the spectrum $\omega_{k}$
with an acoustic law with the velocity $s=\sqrt{\gamma_{11}n/M}$, we
obtain the following equation for $T_c$
\begin{equation}\label{56}
T_c=T_0\left(1-C \frac{T_c^3}{T_0^3}\right),
\end{equation}
where
\begin{equation}\label{57}
   T_0=\frac{\pi}{2}\frac{\hbar^2 n}{M}
\end{equation}
and
\begin{equation}\label{58}
   C=\frac{3
\pi^2\zeta(3)}{16}\frac{\hbar^4}{M^2 \gamma_{11}^2}.
\end{equation}
At $d=a_0$ the interaction constant equals $\gamma_{11}\approx 5 e^2 a_0/\varepsilon= 5 \hbar^2/m$ (see Fig. \ref{f2}).
Accordingly, the constant $C$ is very small ($C < 10^{-2}$) and equation (\ref{56}) with high accuracy gives $T_c= T_0$.
If the excitation spectrum contains a deep minimum (at $d\to d_c$), the critical temperature falls, turning into zero at the instability point.

Let us now estimate how the interaction between pairs and impurities
influences on the transition temperature. As it has been shown in
\cite{46,47}, interaction of Bose particles with short-acting
impurities (with the impurity potential $U_{imp}(\mathbf{R})=\sum_i
U_0\delta(\mathbf{R}-\mathbf{R}_i)$, where $\mathbf{R}_i$ are the
impurity coordinates) leads to decrease of the superfluid density of
the Bose gas $n_s=n_s^0-\Delta n_s^{imp}$ ñ
\begin{equation}\label{59}
    \Delta n_s^{imp}=n_{imp}\frac{M U_0^2}{2\pi \hbar^2 \gamma},
\end{equation}
where $n_{imp}$ is the density of impurities, $M$ is the Bose particle mass and $\gamma$ is the constant of the interaction between the particles which is assumed point-like.
A similar result can be obtained also for the electron-hole system if one adds into the right-hand side of (\ref{23}) a term describing the interaction of pairs with impurities.
This gives an equation for $\Psi(\mathbf{R})$
\begin{eqnarray}\label{60}
\tilde{\mu} \Psi({\bf R}_{12})= -\frac{\hbar^2}{2
M}\nabla^2_{R_{12}}\Psi({\bf R}_{12})+
U_{imp}(\mathbf{R}_{12})\Psi({\bf R}_{12})\cr +\int d {\bf r}_{12} d
{\bf r}_3 d {\bf r}_4 \left[A[{\bf r}_i] \Psi({\bf R}_{12})|\Psi(
{\bf R}_{34})|^2 + B[{\bf r}_i]\Psi( {\bf R}_{32})\Psi^*( {\bf
R}_{34})\Psi( {\bf R}_{14})\right].
\end{eqnarray}
Assuming the interaction with impurities to be weak, we will seek
for a solution of (\ref{60}) in the form
\begin{equation}\label{61}
  \Psi(\mathbf{R})=\Psi_0+\Psi_1(\mathbf{R}),
\end{equation}
where $\Psi_0=\sqrt{n}$.
Substitution of (\ref{61}) into (\ref{60}) gives in the linear approximation the following expression for a Fourier component of the correction $\Psi_1$
\begin{equation}\label{62}
    \Psi_1(\mathbf{q})=\frac{\sqrt{n}}{S [\hbar \omega(q)]^2}
\left[U_{imp}(\mathbf{q})\left(\epsilon_q
    +[\gamma_d(q)+\gamma_{ex}^{(1)}(q)]n\right)
-U_{imp}(\mathbf{-q})[\gamma_d(q)+\gamma_{ex}^{(2)}(q)]n\right].
  \end{equation}
The superfluid density at $T=0$ is determined from the relation \cite{46,47}
\begin{equation}\label{63}
    n_s=n-\frac{1}{2 n}\sum_{\mathbf{q}\ne 0}\langle n_\mathbf{q}
    n_{-\mathbf{q}}\rangle,
\end{equation}
where angle brackets denote averaging by impurity positions and $n_\mathbf{q}=\int d\mathbf{r} e^{-i \mathbf{q}\mathbf{R}}|\Psi(\mathbf{R})|^2$ is the Fourier component of the pair density.
Expressing $n_\mathbf{q}$ in terms of $\Psi_1(\mathbf{q})$, substituting it into (\ref{63}) and calculating the average by impurity positions, we arrive at an expression for the correction to the superfluid density
\begin{equation}\label{64}
    \Delta n_s^{imp}=\frac{ n_{imp}  }{\pi}\int_0^\infty d q
    |U_{\mathbf{q}}^{(0)}|^2
    \frac{n q}{\left(\epsilon_q+2\gamma_q n\right)^2},
   \end{equation}
where
$\gamma_q=\gamma_d(q)+[\gamma_{ex}^{(1)}(q)+\gamma_{ex}^{(2)}(q)]/2$
and $U_{\mathbf{q}}^{(0)}$ is the Fourier component of the potential
of the impurity located in the origin.  The relative change of the
critical temperature can be estimated as $\Delta T_c/T_c=-\Delta
n_s^{imp}/n$. Replacing $\gamma_q$ with the constant
$\gamma=\gamma_{11}$, we obtain the answer (\ref{59}).

In heterostructures with  donor and acceptor layers the dopant atoms
are charged impurities. Usually the dopant layers are located at a
rather large distance $D$ from the conducting layers ($D\gg d$). For
such impurities the Fourier component $U_{\mathbf{q}}^{(0)}$ equals
\begin{equation}\label{65}
    U^{(0)}_\mathbf{q}=(4\pi
e^2/\varepsilon q) \sinh (q d) \exp(-q D).
   \end{equation}
We imply that $d$ is not very close to the critical one and $D$ is
much larger than the healing length $\xi=\hbar\sqrt{M \gamma_{11}
n}$. Substituting (\ref{65}) into (\ref{64}), we obtain
\begin{equation}\label{66}
\frac{\Delta
n_s}{n}\approx\pi\frac{n_{imp}}{n}\frac{M}{m}\left(\frac{e^2
a_0}{\varepsilon\gamma_{11}}\right) \left(\frac{\xi}{D}\right)^2
\left(\frac{d}{a_0}\right)^2.
\end{equation}
For $n_{imp}=2 n$ (the dopant density coincides with the density of
carriers in the conducting layers), $d=a_0$ and $M=4m$ the estimate
(\ref{66}) yields $\Delta n_s/n\approx 5 (\xi/D)^2$. The quantity
obtained is proportional to the square of the small parameter and
under condition $\xi\ll D/\sqrt{5}$ the influence of charged
impurities can be neglected. Note that the latter condition
determines the restriction from below on the density of the pairs.

For estimating the influence of neutral impurities (structure
defects) one can use Eq. (\ref{59}), taking  $U_0= e^2 a$, where $a$
is of order of the lattice parameter. We obtain
\begin{equation}\label{67}
\frac{\Delta n_s}{n}=\frac{\varepsilon^2}{2\pi}
\frac{n_{imp}}{n}\frac{M}{m}\left(\frac{e^2
a_0}{\varepsilon\gamma_{11}}\right)
 \left(\frac{a}{a_0}\right)^2.
\end{equation}
For $M=4 m$, $\varepsilon=13$ and $\gamma_{11}=5 e^2
a_0/\varepsilon$ (that corresponds $d=a_0$) one finds ${\Delta
n_s}/{n}\approx 20 (n_{imp}/n) (a/a_0)^2$. Since $a\ll a_0$ the
condition of smallness of ${\Delta n_s}/{n}$ reduces to the
requirement for the pair density not to be  much less than the
density of neutral defects.

If the distance between the layers is close to the critical one and
the spectrum has a deep minimum, an essential additional
contribution to the integral (\ref{64}) comes from the wave vectors
near the minimum of $\omega(k)$. In this case the expressions given
above underestimate ${\Delta n_s}/{n}$. At $d$ approaching $d_c$ the
negative correction of the critical temperature caused by impurities
will grow up.

It is of interest to compare the influence of impurities on the
superfluidity of the pairs in the systems under study and in quantum
Hall systems \cite{50,51}. The specifics of the latter ones is that
at $d=0$ the gas of electron-hole pairs (magnetoexcitons) is the
ideal one \cite{52}. In that case the expression for the normal
density (\ref{55}) diverges  and the critical temperature goes to
zero. On the other hand, the effective mass of magnetoexcitons grows
up under increase in the interlayer distance, that reduces the
parameter $T_0$ in the equation for the critical temperature
(\ref{56}). It reveals itself in that there exists an optimal $d$ at
which the influence of impurities and other defects will be minimal.
This conclusion was obtained in \cite{50} in the low density limit
$\nu=2\pi \ell_H^2 n\ll 1$. In \cite{51} an analogous result was
obtained for the half-filled Landau level $\nu=1/2$. It was also
shown in \cite{51} that similar to the systems under present study,
in the quantum Hall system with impurities the critical temperature
falls down under approaching the interlayer distance to the critical
one.

\section{Conclusions}

The use of a formalism based on the Keldysh wave function allowed to
determine the region of stability of a superfluid gas of
electron-hole pairs in bilayer systems. The gas of singlet
electron-hole pairs in these systems is two-component. Components
can be distinguished, for example, by the spin of the electron
forming the pair. We have found that at the interlayer distance
$d\gtrsim 0.2 a_0$ ($a_0$ is the effective Bohr radius of the pair)
separation of the system into components will take place. At lower
$d$ a homogeneous mixture of two components will be stable relative
to spatial separation, but in this case instability is expected
relative to formation of a gas of biexcitons. At large interlayer
distances another type of instability develops, namely, instability
related to formation of Wigner crystal-like phase (or a density
wave). The critical distance $d_c$ corresponding to this
instability, enlarges with decreasing the carrier density. At fixed
$d$ the instability occurs at reaching a critical density $n_c$
which is a power-law function of $d$ with a negative exponent. When
increasing the carrier density, the superfluid transition
temperature $T_c$ increases in direct proportion to the density, but
at approaching to $T_c$ it quickly falls down. Interaction with
impurities decreases $T_c$, however, this effect will be significant
only if the concentration of impurities is of the same order or
greater than the density of the pairs.

It follows from the stated above that adjusting the parameters of
the system at which it is possible to obtain the superfluid state of
pairs is a rather delicate problem. The interlayer distance is
limited both from above and from below, furthermore, these limits
can shift with density variation. If the density is decreased, the
interval of allowable $d$ enlarges, but the negative role of
impurities increases too. Nevertheless, based on the results
obtained we consider that it is realistic to achieve rather high
critical temperature. Let us present some estimates. The parameters
that corresponds to AlGaAs heterostructures are
 $m_e=0.067 m_0$, $m_h=0.45 m_0$ and
$\varepsilon=13$ ($m_0$ is the free electron mass). The effective
Bohr radius is $a_0\approx 12$ nm. Under accounting that $m/M\approx
0.11$ (less than $m/M=0.25$ for $m_e=m_h$) the critical density
$n_c$ is approximately in two times smaller than given by
(\ref{51}). Taking $d=0.7 a_0$ and $n=0.4 a_0^{-2}$ one obtain
 the critical temperature $T_c\approx 15$ K. For the system
  MoS$_2$-MoTe$_2$ in the hexagonal BN matrix $m_e=0.47 m_0$,
$m_h=0.62 m_0$ and $\varepsilon=5$. The effective Bohr radius is
$a_0\approx 1$ nm. Due to a small difference of the electron and
hall masses the relation (\ref{51}) is applicable without
correction. Taking $n=0.04a_0^{-2}$ (that corresponds to $d_c\approx
2.3 a_0$) we obtain $T_c\approx 102$ K.

\appendix

\section{General expression for the spectrum}

Here we present general expressions for the functions that enter
into the answer (\ref{44}) for the spectrum. We assume that the
interaction potentials between electrons and holes satisfy the
relation $V_{ee}(r)=V_{hh}(r)$. The sought-for functions are
expressed in terms of Fourier components of the interaction
potentials $V_S(q)=\int d\mathbf{r} V_{ee}(r)e^{-i \mathbf{q}
\mathbf{r}}$, $V_D(q)=\int d\mathbf{r} V_{eh}(r)e^{-i \mathbf{q}
\mathbf{r}}$ and the Fourier component of the bound state wave
function $\phi_q=\int d\mathbf{r} \phi(r)e^{-i \mathbf{q}
\mathbf{r}}$:

\begin{eqnarray}\label{68}
 \gamma^{(d)}_{\bf k}=V_S(k)\int\frac{d^2 p}{(2\pi)^2}\frac{d^2 p'}{(2\pi)^2}
\Big[\phi^*_{\mathbf{p}+\frac{m_h}{M}\mathbf{k}}
\phi_\mathbf{p}\phi^*_{\mathbf{p}'-\frac{m_h}{M}\mathbf{k}}\phi_{\mathbf{p}'}+
\phi^*_{\mathbf{p}-\frac{m_e}{M}\mathbf{k}}\phi_\mathbf{p}
\phi^*_{\mathbf{p}'+\frac{m_e}{M}\mathbf{k}}\phi_{\mathbf{p}'}\Big]\cr+V_D(k)\int\frac{d^2
p}{(2\pi)^2}\frac{d^2 p'}{(2\pi)^2}
\Big[\phi^*_{\mathbf{p}+\frac{m_h}{M}\mathbf{k}}\phi_\mathbf{p}
\phi^*_{\mathbf{p}'+\frac{m_e}{M}\mathbf{k}}\phi_{\mathbf{p}'}+
\phi^*_{\mathbf{p}-\frac{m_e}{M}\mathbf{k}}\phi_\mathbf{p}
\phi^*_{\mathbf{p}'-\frac{m_h}{M}\mathbf{k}}\phi_{\mathbf{p}'}\Big],
\end{eqnarray}

\begin{eqnarray}\label{69}
\gamma^{(1)}_{\bf k}=-\int\frac{d^2 p}{(2\pi)^2}\frac{d^2
 q}{(2\pi)^2} V_S(p)
\Big[\phi^*_{\mathbf{q}} \phi_\mathbf{q}\phi^*_{\mathbf{q}
-\mathbf{p}+\frac{m_e}{M}\mathbf{k}}\phi_{\mathbf{q}-\mathbf{p}+\frac{m_e}{M}\mathbf{k}}+
\phi^*_{\mathbf{q}}\phi_{\mathbf{q}+\mathbf{p}}
\phi^*_{\mathbf{q}+\mathbf{p}+\frac{m_e}{M}\mathbf{k}}\phi_{\mathbf{q}
+\frac{m_e}{M}\mathbf{k}}\cr +
\phi^*_{\mathbf{q}}\phi_{\mathbf{q}-\frac{m_h}{M}\mathbf{k}}
\phi^*_{\mathbf{q}-\mathbf{p}-\frac{m_h}{M}\mathbf{k}}
\phi_{\mathbf{q}-\mathbf{p}}+
\phi^*_{\mathbf{q}}\phi_{\mathbf{q}+\mathbf{p}-\frac{m_h}{M}\mathbf{k}}
\phi^*_{\mathbf{q}+\mathbf{p}-\frac{m_h}{M}\mathbf{k}}\phi_{\mathbf{q}}\cr
- \phi^*_{\mathbf{q}}\phi_{\mathbf{q}}
\phi^*_{\mathbf{q}-\mathbf{p}} \phi_{\mathbf{q}-\mathbf{p}}-
\phi^*_{\mathbf{q}}\phi_{\mathbf{q}+\mathbf{p}}
\phi^*_{\mathbf{q}+\mathbf{p}}\phi_{\mathbf{q} }\Big]\cr +
\frac{1}{2}\int\frac{d^2 p}{(2\pi)^2}\frac{d^2
 q}{(2\pi)^2} V_D(p)
\Big[\phi^*_{\mathbf{q}+\mathbf{p}} \phi_{\mathbf{q}}
\phi^*_{\mathbf{q}+\frac{m_e}{M}\mathbf{k}}
\phi_{\mathbf{q}+\frac{m_e}{M}\mathbf{k}}+ \phi^*_{\mathbf{q}}
\phi_{\mathbf{q}}\phi^*_{\mathbf{q}
+\mathbf{p}+\frac{m_e}{M}\mathbf{k}}\phi_{\mathbf{q}+\frac{m_e}{M}\mathbf{k}}\cr
+ \phi^*_{\mathbf{q}} \phi_{\mathbf{q}}\phi^*_{\mathbf{q}
+\frac{m_e}{M}\mathbf{k}}\phi_{\mathbf{q}-\mathbf{p}+\frac{m_e}{M}\mathbf{k}}
+ \phi^*_{\mathbf{q}} \phi_{\mathbf{q}+\mathbf{p}}\phi^*_{\mathbf{q}
+\frac{m_e}{M}\mathbf{k}}\phi_{\mathbf{q}+\frac{m_e}{M}\mathbf{k}}\cr
+
\phi^*_{\mathbf{q}+\mathbf{p}}\phi_{\mathbf{q}-\frac{m_h}{M}\mathbf{k}}
\phi^*_{\mathbf{q}-\frac{m_h}{M}\mathbf{k}}\phi_{\mathbf{q} }
+\phi^*_{\mathbf{q}}\phi_{\mathbf{q}-\frac{m_h}{M}\mathbf{k}}
\phi^*_{\mathbf{q}+\mathbf{p}-\frac{m_h}{M}\mathbf{k}}\phi_{\mathbf{q}
}\cr + \phi^*_{\mathbf{q}}\phi_{\mathbf{q}-\frac{m_h}{M}\mathbf{k}}
\phi^*_{\mathbf{q}-\frac{m_h}{M}\mathbf{k}}\phi_{\mathbf{q}-\mathbf{p}
}+\phi^*_{\mathbf{q}}\phi_{\mathbf{q}+\mathbf{p}-\frac{m_h}{M}\mathbf{k}}
\phi^*_{\mathbf{q}-\frac{m_h}{M}\mathbf{k}}\phi_{\mathbf{q}}\cr -
\phi^*_{\mathbf{q}+\mathbf{p}}\phi_{\mathbf{q}}
\phi^*_{\mathbf{q}}\phi_{\mathbf{q}
}-\phi^*_{\mathbf{q}}\phi_{\mathbf{q}}
\phi^*_{\mathbf{q}+\mathbf{p}}\phi_{\mathbf{q} } -
\phi^*_{\mathbf{q}}\phi_{\mathbf{q}}
\phi^*_{\mathbf{q}}\phi_{\mathbf{q}-\mathbf{p}}
-\phi^*_{\mathbf{q}}\phi_{\mathbf{q}+\mathbf{p}}
\phi^*_{\mathbf{q}}\phi_{\mathbf{q}}\Big],
\end{eqnarray}

\begin{eqnarray}\label{70}
 \gamma^{(2)}_{\bf k}=-\int\frac{d^2 p}{(2\pi)^2}\frac{d^2
 q}{(2\pi)^2} V_S(p)
\Big[\phi^*_{\mathbf{q}-\frac{m_e}{M}\mathbf{k}}
\phi_{\mathbf{q}-\mathbf{k}}\phi^*_{\mathbf{q}
-\mathbf{p}-\frac{m_h}{M}\mathbf{k}}\phi_{\mathbf{q}-\mathbf{p}}+
\phi^*_{\mathbf{q}-\frac{m_e}{M}\mathbf{k}}\phi_{\mathbf{q}+\mathbf{p}-\mathbf{k}}
\phi^*_{\mathbf{q}+\mathbf{p}-\frac{m_h}{M}\mathbf{k}}\phi_{\mathbf{q}
 }\Big]\cr +
\frac{1}{2}\int\frac{d^2 p}{(2\pi)^2}\frac{d^2
 q}{(2\pi)^2} V_D(p)
\Big[\phi^*_{\mathbf{q}+\mathbf{p}-\frac{m_e}{M}\mathbf{k}}
\phi_{\mathbf{q}-\mathbf{k}}
\phi^*_{\mathbf{q}-\frac{m_h}{M}\mathbf{k}} \phi_{\mathbf{q}}+
\phi^*_{\mathbf{q}-\frac{m_e}{M}\mathbf{k}}
\phi_{\mathbf{q}-\mathbf{k}}\phi^*_{\mathbf{q}
+\mathbf{p}-\frac{m_h}{M}\mathbf{k}}\phi_{\mathbf{q}}\cr +
\phi^*_{\mathbf{q}-\frac{m_e}{M}\mathbf{k}}
\phi_{\mathbf{q}-\mathbf{k}}\phi^*_{\mathbf{q}
-\frac{m_h}{M}\mathbf{k}}\phi_{\mathbf{q}-\mathbf{p}}+
\phi^*_{\mathbf{q}-\frac{m_e}{M}\mathbf{k}}
\phi_{\mathbf{q}+\mathbf{p}-\mathbf{k}}\phi^*_{\mathbf{q}
-\frac{m_h}{M}\mathbf{k}}\phi_{\mathbf{q}}  \Big].
\end{eqnarray}

If the bilayer system is placed in a homogeneous dielectric medium and the dielectric constant
 of the medium $\varepsilon$ coincides with the dielectric constant
 of the interlayer between electron and hole conducting layers,
  and masses of electrons and holes are equal, integrals in (\ref{68}) -- (\ref{70})
  can be written in a more compact form

 \begin{eqnarray}\label{71}
 \gamma^{(d)}_{\bf k}=\frac{4\pi e^2}{\varepsilon k}(1-e^{- k d})\left[\int\frac{d^2 p}{(2\pi)^2}
\phi_{\mathbf{p}} \phi_{\mathbf{p}+\frac{\mathbf{k}}{2}}\right]^2,
\end{eqnarray}
\begin{eqnarray}\label{72}
 \gamma^{(1)}_{\bf k}=-\frac{4\pi e^2}{\varepsilon }
\int\frac{d^2 p}{(2\pi)^2}\frac{d^2
 q}{(2\pi)^2}\frac{1}{p}\Bigg[\Bigg(\phi_\mathbf{q}^2
 \phi_{\mathbf{q}+\mathbf{p}+\frac{\mathbf{k}}{2}}^2+\phi_\mathbf{q}
 \phi_{\mathbf{q}+\mathbf{p}}\phi_{\mathbf{q}+\mathbf{p}+\frac{\mathbf{k}}{2}}
 \phi_{\mathbf{q}+\frac{\mathbf{k}}{2}}-\phi_\mathbf{q}^2
 \phi_{\mathbf{q}+\mathbf{p}}^2\Bigg)\cr -e^{-p d}\left(2\phi^2_\mathbf{q}
\phi_{\mathbf{q}+\frac{\mathbf{k}}{2}}\phi_{\mathbf{q}+\mathbf{p}+\frac{\mathbf{k}}{2}}-\phi^3_\mathbf{q}
\phi_{\mathbf{q}+\mathbf{p}}\right)\Bigg],
\end{eqnarray}

\begin{eqnarray}\label{73}
 \gamma^{(2)}_{\bf k}=-\frac{4\pi e^2}{\varepsilon }
\int\frac{d^2 p}{(2\pi)^2}\frac{d^2
 q}{(2\pi)^2}\frac{1}{p}\Bigg[\phi_\mathbf{q}\phi_{\mathbf{q}+\frac{\mathbf{k}}{2}}
 \phi_{\mathbf{q}+\mathbf{p}+\frac{\mathbf{k}}{2}}
\phi_{\mathbf{q}+\mathbf{p}+\mathbf{k}} -\frac{e^{-p d}}{2}\left(
\phi_\mathbf{q}\phi_{\mathbf{q}+\frac{\mathbf{k}}{2}}
 \phi_{\mathbf{q}+\mathbf{p}+\frac{\mathbf{k}}{2}}
\phi_{\mathbf{q}+\mathbf{k}}+\phi_\mathbf{q}^2\phi_{\mathbf{q}+\frac{\mathbf{k}}{2}}
\phi_{\mathbf{q}+\mathbf{p}-\frac{\mathbf{k}}{2}}\right)\Bigg].
\end{eqnarray}
In (\ref{71}) - (\ref{73}) functions $\phi_\mathbf{q}$ are assumed
real.


\begin{thebibliography}{48}
\bibitem{1} Yu. E. Lozovik, V. I. Yudson, JETP Lett. \textbf{22}, 274 (1975).
\bibitem{2} S. I. Shevchenko, Sov. J. Low
Temp. Phys. \textbf{2}, 251 (1976).
\bibitem{3} Yu. E. Lozovik, V. I. Yudson, Sov. Phys. JETP \textbf{44}, 389 (1976).
\bibitem{4} H. A. Fertig, Phys. Rev. B \textbf{40},  1087 (1989).
\bibitem{5} D. Yoshioka, A.H. MacDonald, J. Phys. Soc. Jpn. \textbf{59}, 4211 (1990).
\bibitem{6} X.G. Wen, A. Zee, Phys. Rev. Lett. \textbf{69}, 1811 (1992).
\bibitem{7} K. Moon, H. Mori, K. Yang, S. M. Girvin, A. H. MacDonald, L. Zheng, D. Yoshioka, S. C. Zhang, Phys. Rev. B \textbf{51}, 5138 (1995).
\bibitem{8} M. Kellogg, J. P. Eisenstein, L. N. Pfeiffer, K. W. West, Phys. Rev. Lett. \textbf{93}, 036801 (2004).
\bibitem{9} E. Tutuc, M. Shayegan, D. A. Huse, Phys. Rev. Lett. \textbf{93}, 036802 (2004).
\bibitem{10} R. D. Wiersma, J. G. S. Lok, S. Kraus, W. Dietsche, K. von Klitzing, D. Schuh, M. Bichler, H.-P. Tranitz, W. Wegscheider, Phys. Rev. Lett. \textbf{93}, 266805 (2004).
\bibitem{11} B. Spielman, J. P. Eisenstein, L. N. Pfeiffer, and K. W. West, Phys. Rev. Lett. \textbf{84}, 5808 (2000).
\bibitem{12} B. Spielman, J. P. Eisenstein, L. N. Pfeiffer, and K. W. West, Phys. Rev. Lett. \textbf{87}, 036803 (2001).
\bibitem{13} D. Nandi, A. D. K. Finck, J. P. Eisenstein, L. N. Pfeiffer, K. W. West, Nature \textbf{488}, 481 (2012).
\bibitem{14} H. Min, R. Bistritzer, J.-J. Su, and A. H. MacDonald, Phys. Rev. B \textbf{78}, 121401(R) (2008).
\bibitem{15} Yu. E. Lozovik and A. A. Sokolik, JETP Lett. \textbf{87}, 55 (2008).
\bibitem{16} B. Seradjeh, H. Weber, and M. Franz, Phys. Rev. Lett. \textbf{101}, 246404 (2008).
\bibitem{17} C. H. Zhang and Y. N. Joglekar, Phys. Rev. B \textbf{77}, 233405 (2008).
\bibitem{18} D. V. Fil and L. Yu. Kravchenko, Low Temp. Phys. \textbf{35}, 712 (2009).
\bibitem{19} M. Y. Kharitonov and K. B. Efetov, Phys. Rev. B \textbf{78}, 241401(R) (2008).
\bibitem{20} M. Y. Kharitonov and K. B. Efetov, Semicond. Sci. Technol. \textbf{25}, 034004 (2010).
\bibitem{21} A. I. Bezuglyj, S. I. Shevchenko, Sov. J. Low Temp. Phys. \textbf{3}, 116 (1977).
\bibitem{22} Yu. E. Lozovik and V. I. Yudson, Solid State Commun. \textbf{21}, 211 (1977).
\bibitem{23} I. Sodemann, D. A. Pesin, and A. H. MacDonald, Phys. Rev. B \textbf{85}, 195136 (2012).
\bibitem{24} Yu. E. Lozovik, S. L. Ogarkov, and A. A. Sokolik, Phys. Rev. B \textbf{86}, 045429 (2012).
\bibitem{25} R. V. Gorbachev, A. K. Geim, M. I. Katsnelson, K. S. Novoselov, T. Tudorovskiy, I. V. Grigorieva, A. H. MacDonald, S. V. Morozov, K. Watanabe, T. Taniguchi, and L. A. Ponomarenko, Nat. Phys. \textbf{8}, 896 (2012).
\bibitem{26} M. P. Mink, H. T. C. Stoof, R. A. Duine, M. Polini, G. Vignale, Phys. Rev. Lett. \textbf{108}, 186402 (2012).
\bibitem{27} A. F. Croxall, K. Das Gupta, C. A. Nicoll, M. Thangaraj, H. E. Beere, I. Farrer, D. A. Ritchie, and M. Pepper, Phys. Rev. Lett. \textbf{101}, 246801 (2008).
\bibitem{28} J. A. Seamons, C. P. Morath, J. L. Reno, and M. P. Lilly, Phys. Rev. Lett. \textbf{102}, 026804 (2009).
\bibitem{29} A. Gamucci, D. Spirito, M. Carrega, B. Karmakar, A. Lombardo, M. Bruna, L. N. Pfeiffer, K. W. West, A. C. Ferrari, M. Polini, and V. Pellegrini, Nat. Commun. \textbf{5}, 5824 (2014).
\bibitem{30} Y. N. Joglekar, and A. H. MacDonald, Phys. Rev. B  \textbf{64}, 155315 (2001).
\bibitem{31} A. R. Champagne,  J. P. Eisenstein,  L. N. Pfeiffer,  K. W. West, Phys. Rev. Lett. \textbf{100}, 096801 (2008).
\bibitem{32} Yu. E. Lozovik, O. L. Berman, JETP \textbf{84}, 1027 (1997).
\bibitem{33} M.Y.J.Tan, N.D.Drummond, R.J.Needs, Phys. Rev. B \textbf{71}, 033303 (2005).
\bibitem{34} C. Schindler, R. Zimmermann, Phys. Rev. B \textbf{78}, 045313 (2008).
\bibitem{35} A. D. Meyerholen, M. M. Fogler, Phys. Rev. B \textbf{78}, 235307 (2008).
\bibitem{36} R. M. Lee, N. D. Drummond, R. J. Needs, Phys. Rev. B \textbf{79}, 125308 (2009).
\bibitem{37} L. V. Keldysh, Coherent States of Excitons, in "Problems of Theoretical Physics", Nauka, Moscow, 1972 (in Russian).
\bibitem{38} J. R. Klauder and B. S. Skagerstam, Coherent States  Applications in Physics and Mathematical Physics, World Scientific, Singapore  (1985).
\bibitem{39a} A. I. Bezuglyi, S. I. Shevchenko,  Phys. Rev. B \textbf{75}, 075322 (2007).
\bibitem{39} A. I. Bezuglyi, S. I. Shevchenko, Low Temp. Phys. \textbf{35}, 373 (2009).
\bibitem{40} S. I. Shevchenko, A. S. Rukin, JETP Letters \textbf{90}, 42 (2009).
\bibitem{41} S. I. Shevchenko, A. S. Rukin, Low Temp. Phys. \textbf{36}, 146 (2010).
\bibitem{42} S. I. Shevchenko, A. S. Rukin, Low Temp. Phys. \textbf{36}, 596 (2010).
\bibitem{43} S. I. Shevchenko, A. S. Rukin, Low Temp. Phys. \textbf{38}, 905 (2012).
\bibitem{44} C. J. Pethick and H. Smith, Bose-Einstein Condensation in Dilute Gases, Cambridge University Press, London  (2002).
\bibitem{md} F.-C. Wu, F. Xue, and A.H. MacDonald, Phys. Rev. B \textbf{92}, 165121
(2015).
\bibitem{45} L. Yu. Kravchenko, D. V. Fil, J. Low Temp. Phys. \textbf{150}, 612 (2008).
\bibitem{46} S. I. Shevchenko, Sov. J. Low Temp. Phys. \textbf{9}, 69 (1983).
\bibitem{47} S. I. Shevchenko, Sov. J. Low Temp. Phys. \textbf{9}, 523 (1983).
\bibitem{50} A. I. Bezuglyi, S. I. Shevchenko, Low Temp. Phys. \textbf{37}, 583 (2011).
\bibitem{51} A. A. Pikalov, D. V. Fil, Nanoscale Research Lett. \textbf{7}, 145 (2012).
\bibitem{52} I. V. Lerner, Yu. E. Lozovik, Sov. Phys. JETP \textbf{53}, 763 (1981).
\end{thebibliography}
\end{document}